\documentclass[reprint,amsmath,amssymb,aps,twocolumn]{revtex4-2}
\usepackage[utf8]{inputenc}
\usepackage{graphicx}
\usepackage{dcolumn}
\usepackage{bm}
\usepackage{color}
\usepackage{natbib}
\usepackage[colorlinks=true,allcolors=blue,urlcolor=black]{hyperref}
\usepackage[normalem]{ulem}
\usepackage[version=3]{mhchem}

\newcommand{\VB}[1]{\mathbf{#1}} % Bold font vector

\let\originalleft\left
\let\originalright\right
\renewcommand{\left}{\mathclose\bgroup\originalleft}
\renewcommand{\right}{\aftergroup\egroup\originalright}

\begin{document}

\title{Engineering of high-\texorpdfstring{$Q$}{Q} states via collective mode coupling in chains of Mie resonators}

\author{%
Mikhail Mikhailovskii\textsuperscript{1}, Maria Poleva\textsuperscript{1}, Nikolay Solodovchenko\textsuperscript{1}, Mikhail Sidorenko\textsuperscript{1}, Zarina Sadrieva\textsuperscript{1}, Mihail Petrov\textsuperscript{1}, Andrey Bogdanov\textsuperscript{1,2}, and Roman Savelev\textsuperscript{1}
}%

\affiliation{%
\textsuperscript{1}School of Physics and Engineering, ITMO University, Saint Petersburg 197101, Russia\\
\textsuperscript{2}International Joint Research Center for Nanophotonics and Metamaterials, Harbin Engineering University, Sansha road 1777, Qingdao 266404, China
}%

\begin{abstract}
Efficient trapping of light in nanostructures is essential for the development of optical devices that are based on the interaction between light and matter.  In this work, we show theoretically and experimentally that one-dimensional arrays of subwavelength dielectric Mie-resonant particles can support collective resonances with increased $Q$-factors. We demonstrate that the increase of the $Q$-factor can be explained by interaction between the collective electric and magnetic dipole modes of the chain resulting in appearance of the inflection point at the band edge. The considered effect is studied experimentally in the chain of high-index ceramic cylinders in the microwave spectral range.
\end{abstract}

\maketitle

\section{Introduction}
Nanostructures that support high-quality (high-$Q$) optical resonances are of great interest for both fundamental research and practical applications. They can significantly enhance the interaction between light and matter, which is one of the key goals in nanophotonics. Recent advances in nanofabrication methods and modelling technologies have led to the development of a rich variety of designs of planar photonic resonators~\cite{KrasnokMetaX2015, KuznetsovScience2016, KivsharSciRev2018, KoshelevNph2019}, including microdisk and ring resonators~\cite{BogaertsLPhRev2012, ZhukovLSA2021}, photonic-crystal (PhC) 
cavities~\cite{EnglundOE2005,QuanOE2011,LaiAPL2014}, arrays supporting bound states in the continuum~\cite{BulgakovPRA2019,SidorenkoPRAppl2021,KoshelevPhysUsp2023} to name a few. The most important characteristic of a photonic resonator, although not universal, is the ratio of its $Q$-factor and the mode volume, which defines the strength of the light-matter interaction. Arguably, the most developed designs in this context are photonic-crystal cavities, showing $Q$-factor up to 10$^6$ with minimal mode volumes for dielectric resonators of the order of $\lambda^3$, where $\lambda$ is the resonant wavelength, and in theory even smaller~\cite{LoncarBook,YangMM2020}.

The recently suggested alternative to the PhC resonators is based on the coupled subwavelength Mie-resonant nanoparticles~\cite{SavelevPRB2014,KrasnokAPL2016,SavelevPRB2017,RutckaiaNanoLett2017, RutckaiaACS2019,HoangNanoLett2020,DingACS2020,DingACS2022}. Due to the interference of radiation from individual nanoparticles, such structures support collective band-edge resonances, characterized by suppressed radiative losses and consequently high $Q$-factors. It has been experimentally shown that such systems allow enhancing the nonlinear optical response~\cite{DingACS2020}, to increase the photoluminescence intensity of the quantum dots integrated into them~\cite{RutckaiaNanoLett2017, RutckaiaACS2019}, to achieve a regime of laser generation~\cite{HoangNanoLett2020}, and act as a platform for on-chip sensing~\cite{DingACS2022}. Although experimental results reported in the literature show relatively moderate $Q$-factors typically growing as $Q \propto N^3$ with the number of the periods in the chain $N$~\cite{ZhangPRL2020}, generally they can have a compact footprint and provide more flexibility owing to the bottom-up approach in their design. Moreover, it was recently predicted that the $Q$-factor of such collective states can be substantially increased in a chain of nanoresonators with a purely dipole response due to coupling between two modes via radiation continuum~\cite{KornovanPRA2019, KornovanACS2021}. However, this requires periods of the chain $a \approx 0.2 \lambda$ (where $\lambda$ is the operational wavelength), which is practically impossible to achieve in realistic optical chains in a resonant regime due to upper limit of the available refractive index $n \lesssim 4$.

\begin{figure}[t]
    \centering \includegraphics[width=1\columnwidth]{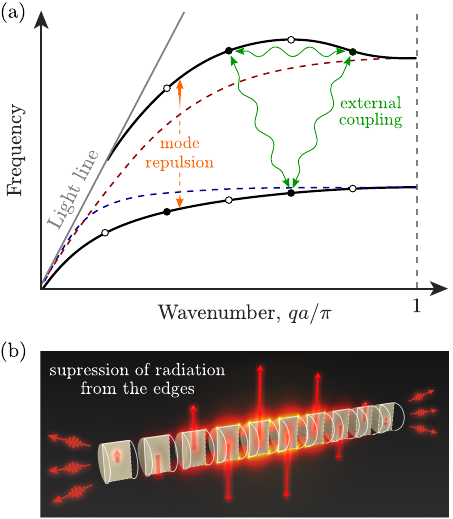}
    \caption{(a) Schematic dispersion of a chain of resonators that supports two uncoupled (dashed curves) and coupled (solid curves) resonances. Solid and hollow circles indicate approximate positions of the even and odd eigenmodes of the corresponding finite chain, respectively. (b)~Schematic electric field distribution in the eigenmode with the highest $Q$-factor; red arrows indicate dipole moments induced in the resonators.}
    \label{fig:schematic}
\end{figure}

On the other hand, all-dielectric Mie nanoresonators exhibit natural multipolar response that can be controlled by changing their geometry~\cite{EvlyukhinPRB2019}. Such control over the response of the individual elements that constitute a bottom-up designed optical structure offers additional degrees of freedom that can produce novel effects. A prominent example is a Huygens scatterer that exhibits electric dipole (ED) and magnetic dipole (MD) responses at close frequencies, which drastically modifies its scattering pattern~\cite{Nieto-VesperinasJOSAa2011,FuNatComm2013}. In the chain of such scatterers, the interaction between MDs and EDs leads to qualitative modification of the chain dispersion, and in particular to the emergence of the backward waves with a locked ratio of magnetic and electric dipole moments~\cite{Shore2012_1,Shore2012_2,StaudeLPR2023}.

Here, we take advantage of the multipolar response of the individual subwavelength resonators for control of the guided waves dispersion in the corresponding one-dimensional arrays. More precisely, we demonstrate that the interplay between the ED and MD resonances in an individual resonator allows switching from typical parabolic dispersion to the non-monotonic one near the band edge of a chain, see Fig.~\ref{fig:schematic}(a). Such modification induces the external radiative coupling of the band edge modes in the finite chain, suppressing radiation from the edges of the chain, Fig.~\ref{fig:schematic}(b). This results in a fast increase of the $Q$-factor as a function of the number of resonators in the chain. Importantly, such an effect can be achieved for relatively large periods of the chain $a \gtrsim 0.3 \lambda$, allowing for a feasible design of optical chain-based resonators. We verify these theoretical predictions with experimental measurements performed for the chain of ceramic cylinders in the microwave spectral range demonstrating a drastic increase in $Q$-factor induced by tuning the distance between the resonators in the chain.

\section{Theoretical model}

\begin{figure*}[t]
    \centering \includegraphics[width=1\textwidth]{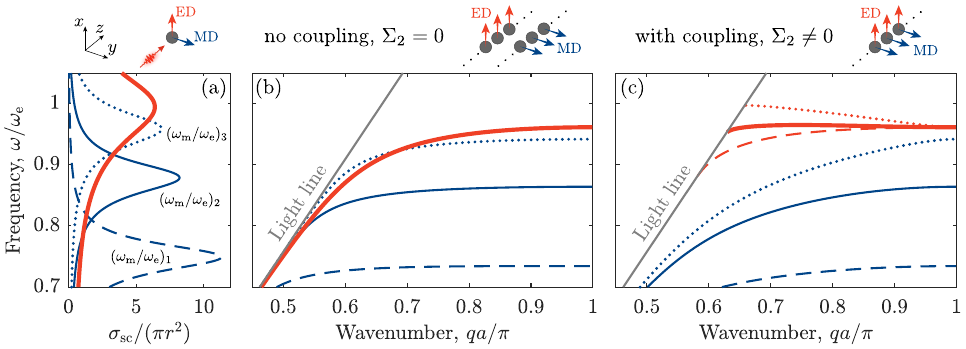}
    \caption{(a) SCSs for a single scatterer with MD (blue curves) and ED (red curve) dipole responses. (b,c) Dispersion diagrams for an infinite chain of dipole scatterers with (b) uncoupled and (c) coupled MD and ED dispersion branches. Red (blue) curves correspond to the modes with dominant ED (MD) contribution. Different linestyles in all panels correspond to different ratios of the MD and ED resonant frequencies: $(\omega_\text{m}/\omega_\text{e})_1 = 0.75$ (dashed curves), $(\omega_\text{m}/\omega_\text{e})_2 = 0.88$ (solid curves), and $(\omega_\text{m}/\omega_\text{e})_3 = 0.96$ (dotted curves).}
    \label{fig:dipole_inf}
\end{figure*}

The conventional coupled-dipole model is a powerful tool that can accurately describe electromagnetic response of single dielectric nanoresonator, the dispersion of an infinite chain, and collective resonances in a finite chain in a certain range of parameters~\cite{novotny_principles_2012,Shore2012_1,Shore2012_2,SavelevPRB2014}. Before proceeding to the full-wave numerical simulations and experimental merasurements, in this Section we employ such a toy model to analyze electromagnetic properties of the infinite and finite chains of dielectric Mie nanoresonators and to gain the understanding the physical mechanisms that determine their characteristics. We model each of the nanoresonators as a combination of magnetic and electric point dipoles with magnetic and electric dipole polarizabilities $\alpha_\text{m,e}$, respectively. The coupling between all EDs and MDs induced in the nanoresonators is described by the electromagnetic Green's tensors~\cite{novotny_principles_2012}. From what follows, we consider only transverse (with respect to the chain axis) orientations of the dipole moments, $p_x$ and $m_y$, see Fig.~\ref{fig:dipole_inf}, since longitudinal polarization does not allow observing the effect under study. Further details on the model are given in Appendix, Sec. A.

At the first step, we analyze which parameters of the infinite chain govern its dispersion properties and how. To this end, we consider the dispersion equation in the following form~\cite{Shore2012_1}: 
\begin{equation}
    \label{eq:DE_coupled}
    \begin{vmatrix} \left( \dfrac{1}{\alpha_\text{m}} - \Sigma_1 \right) & \Sigma_2 \\ \Sigma_2 & \left( \dfrac{1}{\alpha_\text{e}} - \Sigma_1 \right) \end{vmatrix} = 0,
\end{equation}
where $\Sigma_{1,2}(k_0a,qa)$ are the corresponding dipole sums that depend on the free space wavenumber $k_0$, Bloch wavenumber of the mode $q$, and the period of the chain $a$ (see Appendix, Sec. A for the details). When the constituent nanoresonators exhibit dipole response of a single type, either ED or MD, the corresponding equations reduce to:
\begin{equation}
\dfrac{1}{\alpha_\text{m,e}} = \Sigma_1.
\label{eq:DE_uncoupled}
\end{equation}
Therefore, $\Sigma_1$ along with the polarizabilities of the nanoresonator determine the dispersion properties of uncoupled chains of dipoles, while $\Sigma_2$ describes the coupling between two branches.

The strength of the coupling between the MD and ED branches can be controlled mainly by two parameters: (i) the period of the chain $a$, and (ii) the spectral detuning between the uncoupled eigenmodes, with the latter being determined mostly by the MD and ED resonant frequencies of an individual resonator, $\omega_\text{m}$ and $\omega_\text{e}$, respectively. In our model, we employed analytical expressions for the resonant polarizabilities that explicitly contain $\omega_\text{m,e}$:
\begin{equation}
    \alpha_{\text{m,e}} = \dfrac{6\pi}{k_0^3}\dfrac{1}{2Q_{\text{m,e}} \left(1 - \omega/\omega_{\text{m,e}} \right) - \mathrm{i}},
    \label{eq:polarizabilities}
\end{equation}
where $\omega$ is the frequency of the external wave, $k_0=\omega/c$, $c$ is the speed of light, $Q_\text{m,e}$ are the quality factors of the MD and ED resonances, respectively. Although such approximate expressions are not general, they allow to identify the main parameters responsible for the qualitative behaviour of the chain dispersion. In Fig.~\ref{fig:dipole_inf}(a), we plot the scattering cross-sections (SCSs) $\sigma_\text{sc}$ corresponding to electric and magnetic dipoles with polarizabilities given by Eq.~\eqref{eq:polarizabilities} for three different ratios of the resonant frequencies $\omega_\text{m}/\omega_\text{e} = \{0.75,0.88,0.96 \}$. The parameters $Q_\text{m}=13.6$, $Q_\text{e}=6.8$ were chosen in such a way that they approximately correspond to the $Q$-factors of the spherical resonators made of dielectric material with permittivity $\varepsilon=15.15$, which is employed in the further experiments.

The dispersion of the infinite chain of uncoupled [Eq.~(\ref{eq:DE_uncoupled})] and coupled [Eq.~(\ref{eq:DE_coupled})] dipoles are shown in Figs.~\ref{fig:dipole_inf}(b,c), respectively. The values of $\omega_\text{m}/\omega_\text{e}$ are taken as in Fig.~\ref{fig:dipole_inf}(a) and the period is $a=0.33\lambda_\text{e}$. At the edge of the Brillouin zone $qa/\pi=1$, eigenfrequencies of both uncoupled and coupled chains coincide, since $\Sigma_2$ turns to zero at this point, manifesting the different symmetry of the ED and MD standing dipole waves with respect to the chain axis. When shifting away from this point towards the center of the Brillouin zone, the value of $\Sigma_2$ increases, which leads to the repulsion of the dispersion branches. Consequently, the group velocity of the first branch (blue curves) increases, and unlike the case of single-type dipoles, the dispersion curve remains monotonic even for rather small value of the periods $a \approx 0.15 \lambda_\text{m}$, where $\lambda_\text{m}=2\pi c/\omega_\text{m}$ is the resonant wavelength.

At the same time, the group velocity of the second dispersion branch decreases due to repulsion. For $\omega_\text{m}/\omega_\text{e} = 0.75$ it still remains positive, see upper dashed red curve in Fig.~\ref{fig:dipole_inf}(c). However, as $\omega_\text{m}/\omega_\text{e}$ gets closer to $1$, the strength of the repulsion between the branches increases and the group velocity of the upper branch can be pushed to the negative values. When the resonant frequencies are very close to each other $\omega_\text{e} \approxeq \omega_\text{m}$, the $\Sigma_2$ coupling coefficient becomes so large that the upper branch transforms into a backward wave with the quasi-linear dispersion, as illustrated in Fig.~\ref{fig:dipole_inf}(c) with red dotted curve, achieving the so-called ``Hyugens'' regime of the guiding~\cite{StaudeLPR2023}. Consequently, in the intermediate case $\omega_\text{m}/\omega_\text{e}=0.88$, the dispersion of the upper branch is slightly non-monotonic or quasi-flat (i.e. $\mathrm{d}^2\omega/\mathrm{d}q^2$ vanishes for $qa/\pi \rightarrow 1$~\cite{FigotinPRE2005}), as shown with red solid curve in Fig.~\ref{fig:dipole_inf}(c).

\begin{figure}[t]
    \centering \includegraphics[width=1\columnwidth]{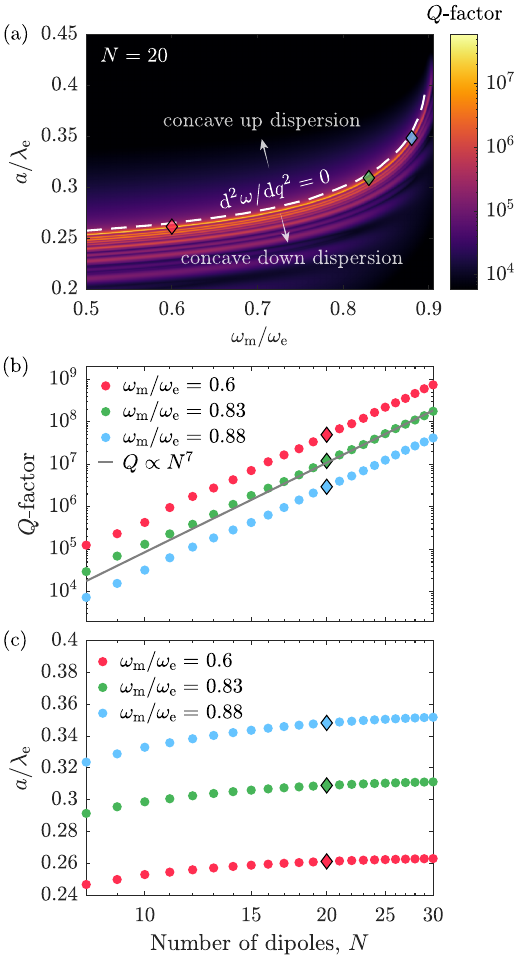}
    \caption{(a) The highest $Q$-factor obtained numerically in the chain of $N=20$ scatterers as a function of the normalized period $a/\lambda_\text{e}$ and the ratio of resonant frequencies $\omega_\text{m}/\omega_\text{e}$. Markers indicate the highest $Q$-factor for several $\omega_\text{m}/\omega_\text{e}$. White dashed line shows the critical period corresponding to zero group velocity dispersion at the edge of the Brillouin zone extracted from the calculations performed for the infinite chain. (b) The highest $Q$-factor and (c) period corresponding to the highest $Q$-factor as a function of the number of the scatterers in the chain $N$ for different $\omega_\text{m}/\omega_\text{e}$. Solid curve in (b) shows the $N^7$ fitting function.}
    \label{fig:finite_dipoles}
\end{figure}

The qualitative change of the dispersion near the edge of the Brillouin zone depends not only on the ratio of the resonant frequencies but also on the period of the structure. In Fig.~\ref{fig:finite_dipoles}(a), white dashed curve shows the set of parameters ($a/\lambda_\text{e}$,\;$\omega_\text{m}/\omega_\text{e}$) for which the dispersion becomes quasi-flat. In the region above the curve, the dispersion is monotonic with positive group velocity, while below the curve it can be non-monotonic or monotonic with negative group velocity. For large spectral separation of the resonances, the critical period approaches $a_{0} \approx 0.24\lambda_\text{e}$, derived previously~\cite{KornovanPRA2019} for the single-type dipole chain, while for the small separation, it approaches the diffraction limit $a \approx 0.5\lambda_\text{e}$. 

Such different dispersion properties below and above the dashed line in Fig.~\ref{fig:finite_dipoles}(a) result in the qualitatively different behaviour of the finite chain eigenmodes in these regions. This is illustrated with a colormap in Fig.~\ref{fig:finite_dipoles}(a) showing the calculated maximal $Q$-factor in the finite chain of $N=20$ dipoles as a function of the ratio of the frequencies and the period of the chain. One can observe that in the upper region, the $Q$-factor is smooth with moderate values. On the other hand, in the lower region, the external coupling between the modes leads to the appearance of the series of maxima with the highest one exceeding the $Q$-factors in the upper region by a couple of orders. The position of the highest $Q$-factor in the ($a/\lambda_\text{e}$,\;$\omega_\text{m}/\omega_\text{e}$) parametric space approaches the dashed curve with the increase of $N$. Note that the calculations presented in Fig.~\ref{fig:finite_dipoles} were performed within the quasi-resonant approximation, i.e. the coupling constants (the Green's functions) were assumed to be frequency-independent and were calculated at the frequency $\omega_\text{e}$.

For each size of the chain $N$, there exists a set of the values ($a/\lambda_\text{e}$,\;$\omega_\text{m}/\omega_\text{e}$) that corresponds to the maximal $Q$ (under variation of one of the parameters). As an example, we choose three values of $\omega_\text{m}/\omega_\text{e}$ and optimize the period of the chain in the range $N=8 \ldots 30$ for each of the $\omega_\text{m}/\omega_\text{e}$. The results are shown in Fig.~\ref{fig:finite_dipoles}(b). First of all, we note that for different values of $\omega_\text{m}/\omega_\text{e}$ the dependence of the $Q$-factor on the size of the chain is approximately $\propto N^7$ in correspondence with the previous results on the single-type dipole chain~\cite{KornovanPRA2019}. The different ratio of resonant frequencies mainly affects the absolute values of the $Q$-factor, while the $N^7$ dependence remains the same. We also notice that as the number of the dipoles increases, the optimal period tends to the asymptotic value, as shown in Fig.~\ref{fig:finite_dipoles}(c). The possibility to achieve the boost in $Q$-factor for rather large periods of the chains is especially crucial in the context of realistic optical designs based on a semiconductor platform. Rather small thickness of the semiconductor layers in commercially available wafers along with restrictions of the industrial fabrication protocols and Mie resonance condition imply the minimal size of the period $a \gtrsim 0.3\lambda$~\cite{BakkerNL2017,DingNanoscale2020,StaudeLPR2023}.

\section{Chain of dielectric cylinders: numerical simulations} \label{sec:ceramic_chain}

The proposed mechanism described within the framework of the toy dipole model can be realized in the chain of high-index dielectric resonators with a suitable choice of their parameters. The spectral positions of the ED and MD resonances can be easily controlled in a non-spherical, e.g. cylindrical, high-index dielectric resonator by changing its aspect ratio. Based on the theoretical estimates we have chosen the following parameters of the ceramic cylinder: radius $r=5$~mm, height $h=7$~mm, and dielectric permittivity $\varepsilon=15.15$. Such commercially available resonators allow us to conduct proof-of-concept experiments in the microwave frequency range.

\begin{figure*}[t]
    \centering \includegraphics[width=1\textwidth]{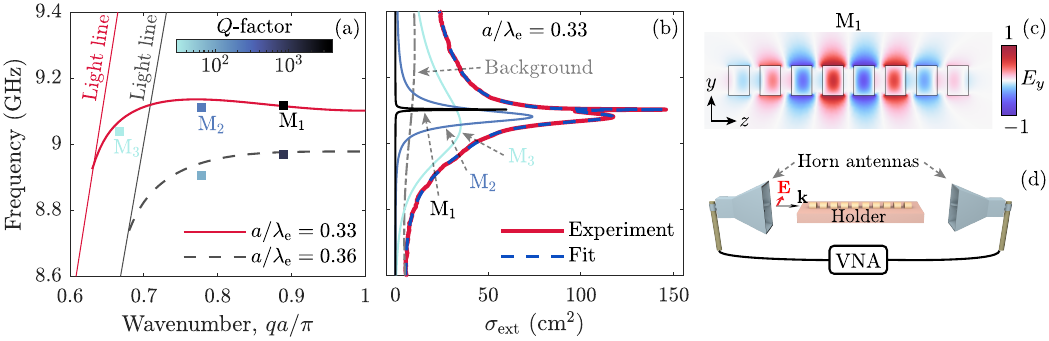}
    \caption{(a) Dispersion diagrams for an infinite chain of ceramic cylinders for the period $a=11.65$~mm (grey dashed curve) and $a=10.6$~mm (solid red curve). Labels M$_1$, M$_2$, M$_3$ indicate the frequencies and quasi-wavenumber of the three eigenmodes in the chain of 8 cylinders; the color of the markers correspond to the $Q$-factor of the eigenfrequencies. (b) Experimentally obtained extinction cross section spectrum for a chain of 8 cylinders measured for the period $a=10.6$~mm (thick red curve). Blue dashed curve shows the fitted spectrum, while thin solid and dashed-dotted curves show contributions from individual modes and the background, respectively. (c) The side-view cross-section of the numerically calculated $E_y$ distribution for the mode M$_1$. (d) Scheme of the experimental setup.}
    \label{fig:numerics}
\end{figure*}

The parameters of the cylinder were chosen in such a way that the ratio of the frequencies corresponding to the maxima of the MD and ED scattering contributions was $\omega_\text{m}/\omega_\text{e} \approx 0.83$ with the $\omega_\text{e} = 2\pi\cdot9.28$~GHz ($\lambda_\text{e}=32.3$~mm), see Appendix, Sec. B. According to the theoretical analysis, the period corresponding to the quasi-flat dispersion in this case is $a \approx 0.31 \lambda_\text{e} \approx 10$~mm. Full-wave numerical simulations of the chain of ceramic cylinders qualitatively confirm this prediction: for the period $a \approx 0.36 \lambda_\text{e} = 11.65$~mm the dispersion is quartic near the edge of the Brillouin zone, dashed grey curve in Fig.~\ref{fig:numerics}(a), while for the smaller periods the dispersion becomes non-monotonic, red curve in Fig.~\ref{fig:numerics}(a). A slight difference between the critical periods is attributed to the simplified expressions for the polarizabilities employed in the theoretical analysis, Eqs.~\eqref{eq:polarizabilities}. Note that the other dispersion branches are well separated in frequency from the considered one, thus simplifying the experimental observation of the resonances in the finite chain, see Appendix, Sec. C. The resonant frequencies of the eigenmodes of the finite chain of $N=8$ cylinders are shown in Fig.~\ref{fig:numerics}(a) with the square markers colored to indicate the $Q$-factor of the corresponding resonances. A particular wavenumber for each mode of a finite chain was determined based on the field distribution of the dominant field component~\cite{KornovanACS2021}: $(qa/\pi)_s = s/(N+1)$, where $s-1$ is the number of sign changes in the distribution of the dipole moments in the chain. For instance, according to the field distribution in Fig.~\ref{fig:numerics}(c) $s=8$ for the mode M$_1$, making it the closest one to the edge of the Brillouin zone. As expected, around the optimal period the finite chain exhibits external coupling of the Fabry--P\'erot resonances, which leads to the $Q$-factor boost of the mode M$_1$ as compared to the same mode in the chain with larger periods.

\section{Experimental verification}

To experimentally verify these results, we have measured the spectral response of the ceramic cylinder chains in the far-field domain under plane wave illumination. The wavevector of the plane wave was directed along the chain axis, thus ensuring the excitation only of the modes with azimuthal numbers $m=\pm1$. The plane wave was generated and the scattered wave was collected by the two horn antennas connected to the ports of a vector network analyzer, as schematically shown in Fig.~\ref{fig:numerics}(d), see additional information in Appendix, Sec. D. The experimentally obtained extinction cross section spectrum (recalculated from the forward scattering via optical theorem) for a chain consisting of 8 cylinders and the optimal distance $a_\text{cr}=10.6$~mm is shown in Fig.~\ref{fig:numerics}(b) with the red solid line. The spectrum clearly exhibits resonances related to the modes M$_1$, M$_2$, M$_3$ indicated in Fig~\ref{fig:numerics}(a). To extract the frequencies and $Q$-factors of these modes we perform the fitting procedure of the extinction spectra, elaborated in Appendix, Sec. E. The result of the fit along with the separate resonant contributions from the eigenmodes are shown in Fig.~\ref{fig:numerics}(b). 

\begin{figure}[t]
    \centering \includegraphics[width=0.9\columnwidth]{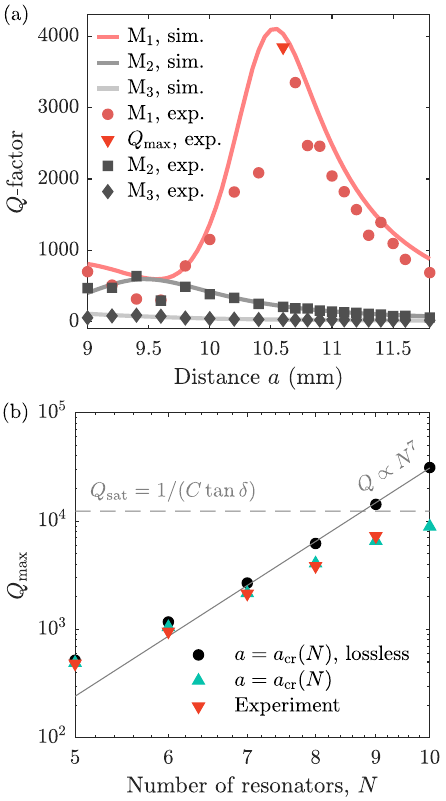}
    \caption{(a) Numerically calculated (solid curves) and extracted from the experiment (markers) dependence of the $Q$-factor on the period in the chain of 8 cylinders. (b) Dependencies of the highest $Q$-factor on the number of cylinders. Black circles correspond to the numerically calculated lossless cylinders, cyan upward-pointing triangles to the numerically calculated cylinders with $\tan\delta = 1.5 \times 10^{-4}$, and red downward-pointing triangles shows the values extracted from the experiment.}
    \label{fig:N}
\end{figure}

Such a procedure allows us to determine the characteristics of individual eigenmodes excited in the chain for different parameters. In Fig.~\ref{fig:N}(a) we show $Q$-factors of the three most weakly radiative modes of the chain composed of 8 cylinders; numerically calculated results are shown with solid curves, while results extracted from the experiments are shown with markers. One can observe that experimental results are in well correspondence with the calculated ones, which confirms the main theoretical predictions. The slight quantitative difference can be explained by the disorder in the geometrical and material parameters of the real structure, such as deviations in the positions of individual cylinder in the chain due to the finite precision of the holder fabrication, as well as variations in the size and dielectric permittivity of individual cylinders, as discussed in Appendix, Sec. F.

Performing the same procedure for different number of resonators in the chain, we obtain the dependence of $Q_\text{max}$ versus $N$, which is shown in Fig.~\ref{fig:N}(b) with red downward-pointing triangles. The corresponding numerically calculated $Q_\text{max}(N)$ dependencies are shown in Fig.~\ref{fig:N}(b) with black circles and cyan upward-pointing triangles for lossless and lossy cases, respectively. The numerical results demonstrate that in the lossless case the $Q$-factor grows approximately as $Q \propto N^7$ in agreement with the theoretical prediction based on a toy dipole model. The material losses, $\tan \delta \equiv \operatorname{Im}\varepsilon/\operatorname{Re}\varepsilon = 1.5 \times 10^{-4}$ for the considered ceramic material, lead to saturation of the $Q$-factor up to $Q_\text{sat} = 1/(C \operatorname{tan}\delta) \approx 1.23\times 10^{4}$, where the prefactor $C$ determines the fraction of electric energy stored inside the resonators. Consequently, for the lossy resonators the $Q_\text{max}(N)$ dependence does not reach $N^7$.

We would like to emphasize that the transverse geometrical dimensions of the considered dielectric chain of resonators are only $\approx0.07 \lambda^2$. Although for large longitudinal size (number of the nanoresonators) the polynomial $N^7$ growth of the collective band-edge mode $Q$-factor loses to the exponential growth in e.g. whispering gallery modes or photonic-crystal nanobeam microresonators, for small sizes their characteristics can be comparable or even better in the chain resonators. For instance, $Q$-factors more than 1000 can be reached in the structures with the longitudinal size $\lesssim 2 \lambda$ (geometrical volume $\approx0.15 \lambda^3$). We also note that although the optimized chain of nanoresonators, i.e. with varied gaps and/or heights of the cylinders, can exhibit higher $Q$-factors than the regular ones, the difference is rather small in the case of the small chains (4-8 nanoresonators)  ~\cite{PichuginNanophotonics2021,PanOL2022,PichuginPhotonics2023,VolkovArxiv2023}. The compactness and geometrical simplicity of the considered regular nanoresonator chain along with the clear physical insight in the physics of the loss suppression might be beneficial for the bottom-up design of the high-$Q$ optical resonant nanostructures.

\section{Summary}
To summarize, we have studied the metamaterial-inspired dielectric cavity based on the chain of subwavelength resonators. We have demonstrated theoretically and experimentally that external coupling of the collective band-edge states in such chain cavities leads to a drastic increase of the $Q$-factor of one of the states. Such effect is linked to the quasi-flat dispersion of an infinite chain, which can be tailored by exploiting the multi-mode electromagnetic response of the resonators. The obtained results have a potential to enrich the variety of the existing approaches in designing compact dielectric resonators for applications in sensing, development of single-photon sources, nanolasers and others. 

\section{Acknowledgements}
The development of the theoretical model was supported by the Federal Academic Leadership Program Priority 2030. The numerical computations and experimental measurements were supported by the Russian Science Foundation, grant No. 22-72-10047. M.M. thanks Alexey Solovyov for useful discussions. N.S. acknowledges support from the Advancement of Theoretical Physics and Mathematics “BASIS” Foundation.

\appendix

\section{Dipole model}

We consider a model of $N$ identical coupled point scatterers that exhibit electric and magnetic dipole responses and are placed at coordinates $\VB{r}_j = ja\VB{e}_z$, $j=1\ldots N$, and $a$ is the period of the chain:
\begin{equation}
    \begin{aligned}
    & \VB{p}_j = \varepsilon_0 \hat{\alpha}_\text{e}\VB{E}(\VB{r}_j),\\
    & \VB{m}_j = \hat{\alpha}_\text{m}\VB{H}(\VB{r}_j),
    \end{aligned}
    \label{eq:dipoles}
\end{equation}
where $\VB{p}_j$ and $\VB{m}_j$ are electric and magnetic dipole moments, respectively, induced in a $j$-th scatterer due to electric $\VB{E}$ and magnetic $\VB{H}$ fields, $\varepsilon_0$ is the vacuum permittivity, and $\hat{\alpha}_\text{e,m}$ are electric and magnetic polarizabilitiy tensors, respectively. Due to symmetry considerations the dipole oscillations can be distinguished by the polarization: two longitudinal --- $p_z$ or $m_z$, and two degenerate transverse ones --- $p_x+m_y$ or $p_y+m_x$. We are interested in the oscillations with transverse orientation of the dipole moments, which allows for coupling between electric and magnetic dipoles. For the scatterers that preserve the axial symmetry of the chain, the polarizability tensors then reduce to the scalar quantities $\alpha_\text{m,e}$. To solve an eigenmode problem we express the field produced at the point of the $i$-th dipole by all other dipoles via the Green's function:
\begin{equation}
\begin{aligned}
    & E_{ix}^\text{ED} = \dfrac{k_0^2}{\varepsilon_0} \sum_{j \neq 0} G_{xx}(\VB{r}_i, \VB{r}_j) p_j,\\
    & E_{ix}^\text{MD} = -\dfrac{\mathrm{i}k_0}{c \varepsilon_0} \sum_{j \neq 0} G_{xy}^\text{EH}(\VB{r}_i, \VB{r}_j) m_j,\\
    & H_{iy}^\text{MD} = k_0^2 \sum_{j \neq 0} G_{yy}(\VB{r}_i, \VB{r}_j)m_j,\\
    & H_{iy}^\text{ED} = -\mathrm{i} k_0 c \sum_{j \neq 0} G_{yx}^\text{HE}(\VB{r}_i, \VB{r}_j) p_j,
\end{aligned}
\label{eq:dipole_fields}
\end{equation}
where $k_0$ is the vacuum wavenumber, $c$ is the speed of light in vacuum, $a$ is the period of the chain and $\widehat{\VB{G}}^\text{HE}(\VB{r}_i,\VB{r}_j) = -\widehat{\VB{G}}^\text{EH}(\VB{r}_i,\VB{r}_j) \equiv \nabla \times \widehat{\VB{G}}(\VB{r}_i, \VB{r}_j)$, where spatial derivative is taken with respect to the first variable $\VB{r}_i$. The Green's function is determined as follows~\cite{novotny_principles_2012}:
\begin{equation}
    \begin{split}
        \widehat{\VB{G}}(\VB{r}_i,\VB{r}_j) &= \dfrac{\mathrm{e}^{\mathrm{i} k_0 R}}{4\pi R}  \left[ \left( 1 - \dfrac{\mathrm{i} k_0 R - 1}{k0^2 R^2} \right) \hat{\VB{I}} \right. \\
        &\mathrel{\phantom{=}} + \left. \dfrac{3 - 3 \mathrm{i} k_0 R - k^2R^2}{k_0^2 R^2} \dfrac{\VB{R} \otimes \VB{R}}{R^2} \right],
    \end{split}
\end{equation}
where $R \equiv |\VB{R}| = |\VB{r}_i - \VB{r}_j|$, and $\hat{\VB{I}}$ is the $3 \times 3$ identity matrix.

For an infinite chain one can apply the Bloch theorem:
\begin{equation}
p_j = p_0 \mathrm{e}^{\mathrm{i}q_j},\;\; m_j = m_0 \mathrm{e}^{\mathrm{i}q_j},
\label{eq:Bloch}
\end{equation}
where $q_j = jqa$, $j$ is an integer number, $q$ is the Bloch wavenumber. Substituting Eqs.~\eqref{eq:Bloch} into Eqs.~\eqref{eq:dipoles} and taking into account expressions for the fields, Eq.~\eqref{eq:dipole_fields}, one obtains the dispersion equation for an infinite chain ({Eq.~(1) in the main text):
\begin{equation}
    \left( \dfrac{1}{\alpha_\text{m}} - \Sigma_1 \right) \cdot \left( \dfrac{1}{\alpha_\text{e}} - \Sigma_1 \right) - \Sigma_2^2 = 0,
    \label{eq:disp_equation}
\end{equation}
where dipole sums $\Sigma_1$ and $\Sigma_2$ are
\begin{gather}
    \Sigma_1 = \frac{k_0^3}{4\pi} \sum_{j \neq 0} \mathrm{e}^{\mathrm{i} (k_j + q_j)} \left( \frac{1}{k_j} + \frac{\mathrm{i}}{k_j^2} - \frac{1}{k_j^3} \right),\\
    \Sigma_2 = \frac{k_0^3}{4\pi} \sum_{j \neq 0} \operatorname{sgn}(j) \mathrm{e}^{\mathrm{i} (k_j + q_j)} \left( \frac{1}{k_j} + \frac{\mathrm{i}}{k_j^2} \right),
\end{gather}
with $k_j = |j|k_0a$. For $q>k_0$ (below the light line) the imaginary part of the $\Sigma_1$ dipole sum $\operatorname{Im}\left( \Sigma_1 \right) = -k_0^3/(6\pi)$ is canceled out with the imaginary part of the inverse dipole polarizability (radiation correction). Along with the zero imaginary part for the cross-coupling term $\operatorname{Im}\left( \Sigma_2 \right) = 0$ this makes dispersion Eq.~\eqref{eq:disp_equation} real-valued, allowing for guided mode solutions.

In the case of finite chains, eigenmodes are found from the matrix equation:
\begin{equation}
    \VB{M}\VB{d}_\text{me}=0,
\end{equation}
where $\VB{d}_\text{me} = (m_1,m_2,\ldots,m_N,p_1,p_2,\ldots,p_N)^{\operatorname{T}}$ is the vector of $2N$ dipole moments and $\VB{M}$ is $2N \times 2N$ interaction matrix consisted of the Green's functions. The $Q$-factor is then determined from the complex eigenfrequency $\omega$ as $Q=-\operatorname{Re}(\omega)/[2\operatorname{Im}(\omega)]$.

The MD and ED contributions to the total scattering cross-section of a single scatterer (Fig.~2(a) in the main text) are calculated as follows:
\begin{equation}
\sigma_\text{sc}^\text{MD,ED} = \dfrac{k_0^4}{6\pi}|\alpha_\text{m,e}|^2.
\label{eq:SCS}
\end{equation}

\section{Properties of a single ceramic resonator}

The desired electromagnetic response of a scatterer, i.e. presence of ED and MD resonant responses with a specific ratio of the resonant frequencies, can be achieved in a simple cylindrical high-index subwavelength resonator. To characterize the response of such particles we have employed the method of multipole analysis of the scattered fields induced in the resonator under the plane wave illumination~\cite{EvlyukhinPRB2019,AlaeeOptComm2018}. Since we are interested in the ED and MD moments oriented along the diameter of the cylinder, the plane wave was incident along the axis of the cylinder. In order to identify the frequencies of the ED and MD resonances we have calculated the partial scattering cross-sections using Eqs.~\eqref{eq:SCS}. The polarizabilities were extracted from Eqs.~\eqref{eq:dipoles}, where ED and MD moments induced in the cylinder by the incident plane wave were calculated via the induced current density $\VB{j}(\VB{r})$:
\begin{gather}
    \begin{split}
        \VB{p} &= \frac{\mathrm{i}}{\omega} \int j_0(k_0\VB{r})\VB{j}(\VB{r}) \, \mathrm{d^3}\VB{r} \\
        &\mathrel{\phantom{=}} + \frac{\mathrm{i}k_0^2}{2\omega} \int \frac{j_2(k_0\VB{r})}{(k_0\VB{r})^2} [ 3(\VB{r} \cdot \VB{j}(\VB{r})) \VB{r} - r^2\VB{j}(\VB{r}) ] \, \mathrm{d^3}\VB{r},
    \end{split}
    \\
    \VB{m} = \frac{3}{2} \int \frac{j_1(k_0\VB{r})}{k_0\VB{r}} [ \VB{r} \times \VB{j}(\VB{r}) ] \, \mathrm{d^3}\VB{r},
\end{gather}
where $j_l(k_0\VB{r})$ is the $l$-th order spherical Bessel function.

\begin{figure}[t]
    \centering \includegraphics[width=0.9\columnwidth]{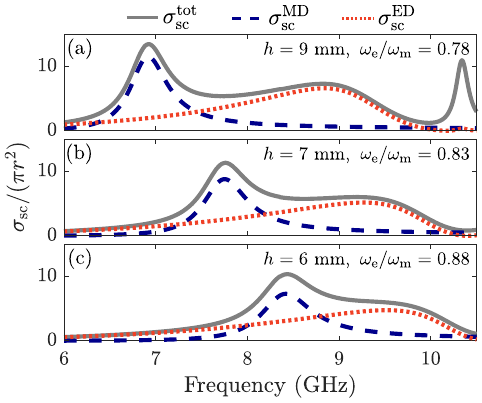}
    \caption{Spectra of SCS (solid grey curves) for a single cylinder with $\varepsilon=15.15$, $r=5$~mm, (a) $h=6$~mm, (b) $h=7$~mm, (c) $h=9$~mm. MD and ED contributions to the SCS are shown with dashed blue and dotted red curves, respectively.}
    \label{fig:multipole}
\end{figure}

The results of the calculations for a ceramic cylinder with permittivity $\varepsilon=15.15$, radius $r=5$~mm, and three different heights $h=6$~mm, $h=7$~mm and $h=9$~mm are shown in Fig.~\ref{fig:multipole}. One can observe that varying the aspect ratio of the cylinder allows one to control the ratio of the MD and ED resonant frequencies in the range $\approx 0.75 \ldots 1$.

\section{Infinite chain of ceramic cylinders}

\begin{figure}[b]
    \centering \includegraphics[width=0.9\columnwidth]{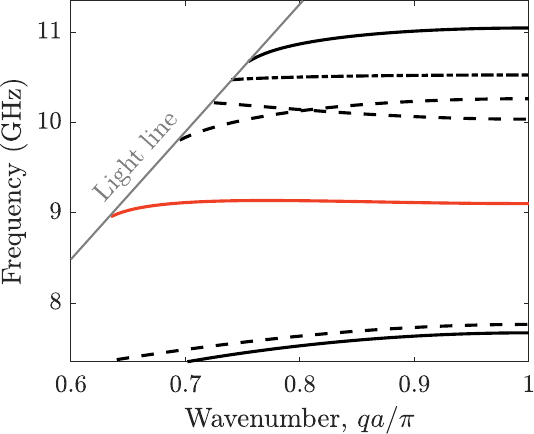}
    \caption{Dispersion for an infinite chain of ceramic cylinders with parameters $h=7$~mm, $r=5$~mm, $\varepsilon=15.15$, $a=10.6$~mm. The dashed curves correspond to the azimuthal number $m=0$, solid curves --- to $m=1$, and the dashed-dotted curve --- to the $m=2$. The non-monotonic mode with dominant ED contribution is shown with the red solid curve.}
    \label{fig:disp_app}
\end{figure}

The dispersion of the first seven guided modes of the infinite chain of coaxially arranged ceramic cylinders with permittivity $\varepsilon=15.15$, radius $r=5$~mm, and height $h=7$~mm are shown in Fig.~\ref{fig:disp_app}. The period of the chain is $a=10.6$~mm. Different styles of the curves correspond to different azimuthal numbers $m$ (see caption in Fig.~\ref{fig:disp_app}). The mode with a dominant ED contribution and azimuthal number $m=\pm 1$ considered in the main text [see Fig.~\ref{fig:numerics}(a)] is shown in Fig.~\ref{fig:disp_app} with red color. One can observe that below the light line the chain of cylinders support a single guided branch over a wide frequency range ($\approx2$~GHz). Consequently, the scattering spectrum of the corresponding finite chain includes contribution from the high-$Q$ collective states associated with this particular branch, while the resonances associated with other branches may contribute only in the form of the non-resonant background. 

\section{Experimental measurements} \label{sec:app. exp}

In experiment we have measured the spectral response of a chain of ceramic cylinders in the far-field domain under illumination by a plane-wave. The excitation and scattered signal were generated and collected by two horn antennas [see Fig.~\ref{fig:numerics}(d)]. 

For precise control of the distance between the individual cylinders in the chain, we fabricated a foam material holder using a computer numeric control machine drilling. The dielectric permittivity of the foam was approximately equal to 1 in the operational spectral range, therefore azimuthal symmetry of the chain was conserved in the first approximation. In all measurements, the direction of the incident plane wave was chosen along the chain axis with polarization of the electric field parallel to the holder-air interface, allowing for an excitation of the superposition of the ED modes with azimuthal number $m=\pm 1$. After measuring the forward scattering signal the extinction cross section spectrum was obtained by applying the optical theorem:
\begin{equation}
    \sigma_\text{ext} = -\dfrac{4 \pi c}{\omega}\operatorname{Im}\left(\dfrac{S_{21}}{S_{21}^\text{bckg}}\right),
\end{equation}
where $c$ is the speed of light, $\omega$ is the angular frequency of the incoming signal, $S_{21}$ is the measured trasmission spectrum of the resonant chain, $S_{21}^\text{bckg}$ is the measured transmission spectrum in the absence of the resonant chain.

\section{Post-processing of experimental data} \label{sec:app. fitting}

Since in the spectra of the extinction cross section several collective modes contribute significantly even in a rather narrow frequency range, the function of the sum of several Fano formulas with an additional polynomial term to account for the background signal was used for further data processing:
\begin{equation}
    f(\omega) = \sum_i^{N_\text{res}} A_i\frac{(q_i+\Omega_i)^2}{1+\Omega_i^2} + g(\omega). \label{formula:Fano}
\end{equation}
Here, $A_i$ is the amplitude of the corresponding $i$-th resonance, $\Omega_i = 2(\omega - \omega_{i})/\gamma_i$, where $\omega$ is the angular frequency of the signal, $\omega_{0,i}$, $\gamma_i$, $q_i$ are the frequency, the width and the Fano parameter of the corresponding $i$-th resonance, respectively, and $g(\omega) = c_1\omega + c_2\omega^2$ is the polynomial term with constants $c_1$ and $c_2$. The number of resonances within the spectral region of interest $N_\text{res}$ was determined from numerical simulations. The parameters of the fitting curves were found by employing an algorithm based on the nonlinear least squared method. The eigenfrequencies obtained from numerical simulations were chosen as initial values of $\omega_{0,i}$ and $\gamma_{0,i}$, while the final values were used to calculate the resulting $Q$-factor of the $i$-th resonance as $\omega_{i}/\gamma_i$. The described fitting procedure was tested on the numerically simulated extinction cross section spectra with additional random noise and provided robust results.

\section{Estimating the influence of disorder}

\begin{figure}[t!]
    \centering \includegraphics[width=1\columnwidth]{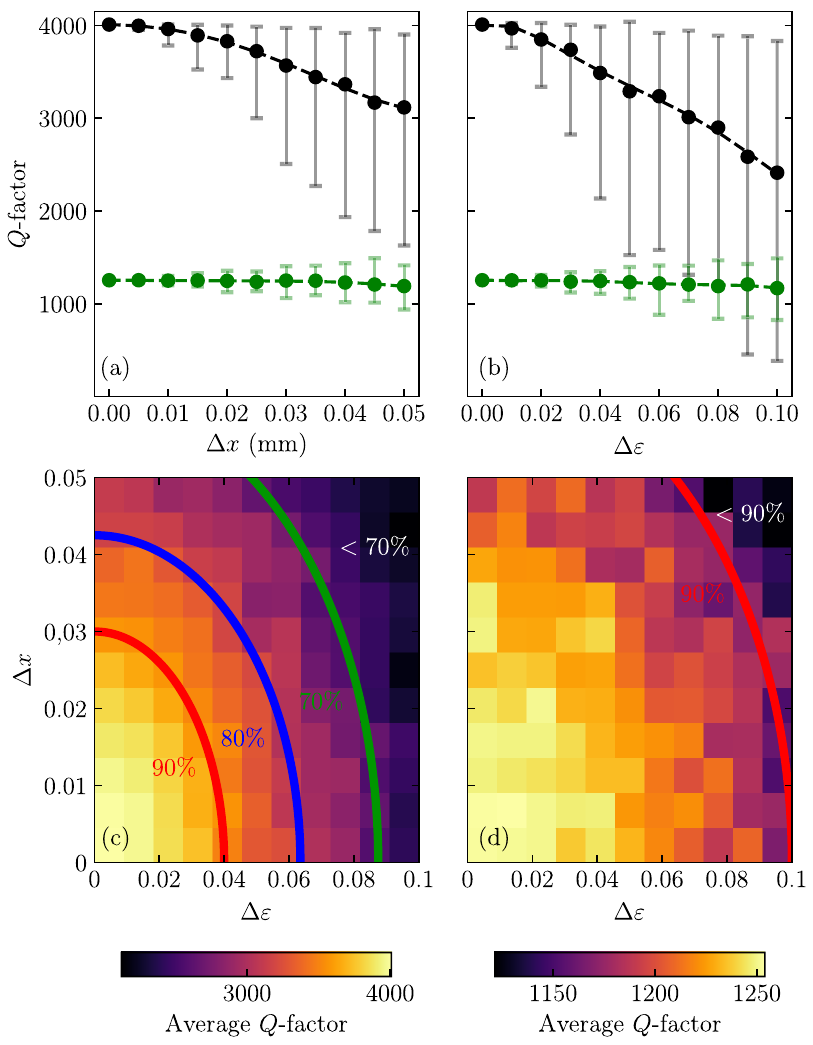}
    \caption{(a,b) Dependence of the highest $Q$-factor for the eigenmodes of the 8 cylinder chain on the standard deviation of (a) longitudinal shift of the cylinders and (b) dielectric permittivity of the ceramic material. The black and green points correspond to the values averaged over 100 calculations for the periods $a_\text{cr}=10.6$~mm and $a=11.5$~mm, respectively; vertical lines indicate the variation of the $Q$-factors. (c,d) $Q$-factor averaged over 100 calculations with simultaneous random deviations in the positions and dielectric permittivities in the chain of 8 cylinders with the periods $a_\text{cr}=10.6$~mm and $a=11.5$~mm, respectively.}
    \label{fig:a-disorder}
\end{figure}

In the experimentally studied chain of ceramic cylinders variation in the material parameters of different cylinders as well as deviations in their positions are inevitably present. To estimate the influence of different types of disorder on the $Q$-factors of the eigenmodes we performed a series of numerical simulations for structures with normally distributed random variations in positions of the cylinders along the chain axis or/and variations in their permittivity. The standard deviation of the corresponding distributions was denoted as $\Delta x$ and $\Delta \varepsilon$, respectively. The results of simulations are presented in the Fig.~\ref{fig:a-disorder}. The averaged highest $Q$-factor for the chain of 8 cylinders with the optimal period $a_\text{cr}=10.6$~mm and larger period $a=11.5$~mm are shown with black and green dots, respectively. The vertical bars show the variation of the $Q$-factors for different random realizations. Figs.~\ref{fig:a-disorder}(a,b) correspond to the disorder in position or permittivity, respectively, and Figs.~\ref{fig:a-disorder}(c,d) -- to simultaneous disorder in position and permittivity.

In the experiments the deviation of the measured $Q$-factor from the simulated one was on average $\approx 10\%$ and maximum $\approx 50\%$ for the periods close to optimal, see Fig.~\ref{fig:N}(a) in the main text. One can observe that for the optimal period $a = 10.6$~mm $50\%$ drop in $Q$-factor requires rather strong average disorder. For a single random realization, however, it can occur for smaller disorder $\Delta x \approx 0.03$~mm or $\Delta \varepsilon \approx 0.04$. Such deviations are less than the ones estimated from the accuracy of the experimental setup and the electromagnetic properties of the fabricated samples.

For the larger period $a=11.5$~mm even strong disorder on average results in the moderate decrease of the $Q$-factor. While for a given random realization the $Q$-factor can even increase by a noticeable amount. This is not surprising, since even random decrease of the distances between the cylinders, which makes them closer to the optimal one, can lead to the increase of the $Q$-factor. Overall, the results of these simulations qualitatively explain the difference between the $Q$-factors obtained in the simulation and extracted from the experiment.

\bibliography{refs}

\end{document}